\documentstyle[manuscript,aps,epsf]{revtex}

\begin{document}
\def\br{{\bf r}}
\def\fy{{\vec\phi}}
\def\valpha{{\vec\alpha}}
\def\vbeta{{\vec\beta}}
\draft
%\bibliographystyle{unsrt}
%\hfil{\today}
\title{ Probability distributions for one component equations
with multiplicative noise.}
\author{J.M. Deutsch}
\address{University of California\\
Santa Cruz CA 95064\\ }
\maketitle
%\vskip 1.5cm
\begin{abstract}
Systems described by equations involving both multiplicative and additive noise are common
in nature. Examples include convection of a passive scalar field, polymers
in turbulent flow, and noise in dye lasers. 
In this paper the one component version of this problem is studied.
The steady state probability distribution is classified into
two different types of behavior. 
One class has power law tails and the other is of the form of
an exponential to a power law.
The value of the power law exponent is determined
analytically for models having colored
gaussian noise. It is found to only depend on the
power spectrum of the noise at zero frequency. When non-gaussian
noise is considered it is shown that stretched exponential tails
are possible. An intuitive understanding of the results is found and
makes use of the Lyapunov exponents for these systems.
\end{abstract}
\newpage
\narrowtext
\section{INTRODUCTION}
\par
This paper analyzes linear equations containing additive and
multiplicative noise. 
There are many problems in physics which are in this category.
The two types of noise are generally uncorrelated with
each other. One example is a polymer in a turbulent
flow. Analysis of this system yields an equation for bending modes of
the polymer with two types of noise~\cite{armstrong,deutsch}. 
Thermal noise presents itself
in an additive way, while the effects of the random velocity field
come in multiplying the coordinates of the polymer, and is hence
termed multiplicative. The random flow field, being turbulent, has
correlations with long decay times, and are therefore highly colored.
Other examples include motion of a passive scalar in a random
velocity field~\cite{pope,eswaran}, and propagation of light in 
random media~\cite{fock}. Noise in dye lasers is modeled by multiplicative
random equations~\cite{zhu}.
Most of the applications involve more than one degree of freedom,
however the multi-dimensional case is surprisingly similar
to the one component case, so it is worth studying it first.
The multi-component case is considered in a separate publication~\cite{many}.

Because of its physical applications, it seems worthwhile to
attempt to understand general properties of such equations.
Drummond~\cite{drummond} recently analyzed such a linear one component
equation of the type of interest in this paper. He considered the
case where both the additive and multiplicative noise where generated
by the same random process. He found two types of behavior are possible.
In one regime he found that the probability distribution  function (PDF)
had power
law tails, so that high enough moments of the variable of interest
did not exist. The other regime had all its moments defined.

In the applications mentioned above, the additive and the
multiplicative noise are not correlated with each other which is
unlike the cases considered by Drummond~\cite{drummond}. Fortunately
this problem is also tractable analytically and is the subject of
this paper. In this case one can go considerably further in
classifying the type of behavior possible, and intuitively how it
occurs. 

There is a large mathematical literature on 
the stability of the nth moment of a variable in the absence of
additive noise which  has been studied by the introduction of Lyapunov
functions~\cite{arnold}. 
The stability of moments can be determined if such a function 
can be found. The method can be extended to additive noise
but does not provide a classification of the kinds of behavior expected
for the tails of the distribution.

The classification instead hinges on an important conceptual framework,
the Lyapunov {\it exponents} for these systems without additive noise,
as discussed in section \ref{sec:absence}. 
All the behavior found here is easily understood in this framework.
In this case the PDF does not tend to a time independent
limit, however the general scaling form of the distribution has been
previously obtained~\cite{benzi,paladin} and allows one to
define a set of Lyapunov exponents $L(q)$.
These are not related simply to Lyapunov functions~\cite{arnold}.
The $L(q)$'s can then be used to give a more intuitive
derivation of the form of the probability distribution's 
tail in the presence of additive noise
discussed in the sections that follow.

The classification into different two regimes is now briefly
described. In one regime, it is possible
to show that the probability distribution of fields satisfying
these types of equations have {\it power law tails}. 
This class of problems occur under a wide variety
of conditions. This will be seen first in section \ref{sec:heuristic} 
where a heuristic argument is given to understand when such tails
are present and when they should be absent.
Then, in section ~\ref{sec:gaussian}, the one dimensional 
version of this problem with gaussian noise is analyzed.
In general these power law tails will be present
with Gaussian noise, even if it is not short range in
time. In this case it is possible to find the exact exponent for the
asymptotic probability distribution, for arbitrary time correlations
in the multiplicative noise. The exponent depends continuously
on the strength of the multiplicative noise in a simple way,
only through the power spectrum at zero frequency. 

The other type of behavior is where all moments are defined and
the PDF has a tail falling faster
than any power law. This is shown in section \ref{sec:nongaussian}
to occur for one dimensional
models but requires that the multiplicative noise {\it not be Gaussian}.
In this case the PDF will be of the form of an exponential to
a power law, the precise exponent depending on the Lyapunov
exponents for high moments. This can result in a stretched
exponential distribution.

\section{One component multiplicative equation}
\label{sec:multi}
\subsection{General equation for the moments}
\label{sec:gen}
\par
Here we will consider the equation
\begin{equation}
 \nu {dx \over dt} ~=~ (-k+ f(t)) x + \xi (t) 
\end{equation}
This can be interpreted as the equation for a massless particle
with friction coefficient $\nu$ connected to a spring that has
a spring coefficient that varies randomly in time as $k - f(t)$.
The spring is perturbed by a random force $ \xi (t) $ representing
a heat bath at temperature T. As a result the $\xi (t) $ is
Gaussian and $\delta$ function correlated $\langle \xi (t) \xi (t') \rangle ~=~
2 \nu T \delta (t-t') $ ~\cite{landau}.
$f(t)$ in general can be considered non-Gaussian with a correlation function
\begin{equation}
\langle f(t) f(t')\rangle ~=~ g(t-t'),~~~\langle f(t)\rangle  ~=~ 0 
\end{equation}
Writing $\tau ~=~ (k/\nu ) t$,
this can be rewritten as 
\begin{equation} {dx \over d \tau} ~=~ -x + \gamma (\tau) x + \eta (\tau) 
\label{eq:multiplicative}
\end{equation}
with 
\begin{equation}
\langle \eta(\tau ) \eta(\tau ') \rangle  ~=~ {2T\over k} \delta (\tau -\tau ')
\end{equation}
and 
\begin{equation}
\langle \gamma (\tau)\gamma (\tau ')\rangle  ~=~ {1\over k^2} g( {\nu \over k} (\tau - \tau ')).
\equiv \sigma (\tau -\tau ' )
\end{equation}
\par

\section{distribution in the absence of additive noise}
\label{sec:absence}
Here we will consider eqn. (\ref{eq:multiplicative}) when the additive noise
term $\eta $ is zero. In this case there is not a well defined
steady state probability distribution. However it is useful to understand
the time dependent form of the distribution for long times.
We shall see that in this limit it has a scaling form.
The main results found in this section will survive the leap to many components and provide
an important conceptual framework for understanding these systems.

First we review why it is that the behavior of the system is 
characterized by Lyapunov exponents~\cite{benzi,paladin} 
\begin{equation}
\langle x(t)^q\rangle \propto e^{L(q)t}
\label{eq:lyapunov}
\end{equation}
where the brackets denote an ensemble average over the noise $\bf \gamma$.
To understand this, 
it is simplest to dispense with the linear term $-kx$  in
eqn. (\ref{eq:multiplicative}) at the expense of 
giving $\gamma(t)$ a nonzero mean $\langle \gamma 
\rangle ~=~ -1$.  A straightforward
way of defining a short range non-gaussian process is to discretize
the above  eqn. (\ref{eq:multiplicative}), $t = i\Delta t$ 
and using It{\^ o} discretization, this equation with $\eta~=~0$ reads
\begin{equation}
x_{i+1}~=~x_i (1+\Delta t \gamma_i) 
\label{eq:discrete}
\end{equation}
This defines a multiplicative process. If the different $\gamma$'s
are all  taken to be independent the Lyapunov exponents can 
be determined by calculating
\begin{equation}
\langle {x_i}^q\rangle ~=~ \langle (1+\Delta t\gamma_1)^q\rangle~=~
e^{L(q)t}
\end{equation}
Hence the Lyapunov exponents are $L(q) ~=~ 
(1/\Delta t)\ln(\langle(1+\Delta t\gamma)^q\rangle )$. In general
these are not quadratic and depend on the probability distribution of
the $\gamma_i$'s. $L(q)$ should be convex. These multiplicative
random processes have been well studied, particularly in recent
years in connection with ``multifractals''.

Given the moments of $x$ it is possible to calculate
its PDF for long times by performing a steepest descent
analysis identical the case of multifractals~\cite{benzi}. It 
can be checked that the distribution giving such scaling is
\begin{equation}
\ln P(\ln x) \propto t f((\ln x) /t) + {\cal O}({\ln t \over t} )
\label{eq:plnphi}
\end{equation}
where 
\begin{equation}
f(\alpha ) = L(q) - q\alpha ,~~~~~~~  \alpha = L'(q).
\label{eq:falpha}
\end{equation}

From this $\langle (\ln x )^2\rangle$ can be calculated and is
proportional to $t$ for large $t$, so the the distribution 
continues to broaden.  Therefore $P (x)$ does not tend towards
a time independent, that is steady state, distribution.

Do these results persist when there are correlations between the different $\gamma_i$?
The answer is that if the correlations are not too long range, these results are
still valid. This can be seen by making an analogy with a statistical mechanical
model. Denote the random variable $1+\Delta t \gamma_i$ in eqn. (\ref{eq:discrete})
by $\exp (z_i)$. Then given the joint PDF of the variables $\{z_1,\dots ,z_n\}$
\begin{equation}
P\{z_1,\dots ,z_n\} \equiv e^{-V\{z_1,\dots ,z_n\}}
\end{equation}
$\langle{x_n}^q\rangle$ can be expressed as
\begin{equation}
\langle{x_n}^q\rangle ~=~ \int \prod_{i=1}^n e^{\sum_{i=1}^n (q-1)z_i
-V\{z_1,\dots ,z_n\}} dz_1\dots dz_n
\end{equation}
This expression can be thought of as the partition function of a system
with $n$ degrees of freedom. For large $n$ the free energy of this is
extensive if $V$ is sufficiently short range~\cite{lieb}. This means that for large
$n$ $\langle{x_n}^q\rangle\sim \exp ({n\Delta t L(q)})$ 
where $L(q)$ can be interpreted
as the free energy per degree of freedom.

\section{Heuristic explanation of power law tails}
\label{sec:heuristic}

Before launching into a detailed mathematical analysis of this
problem, it is worthwhile giving a heuristic explanation for
the existence of power-law tails and their relationship
to the Lyapunov exponents $L(q)$. 

In the absence of additive noise, we just saw that eqn. (\ref{eq:multiplicative})
can be characterized by the different moments of $x_i$. If we examine
the qth moment $\langle x_i^q\rangle$, it will either 
exponentially  increase or decrease depending on whether $L(q)$ is
positive or negative respectively. 

Now examine how the last statement changes in the presence of
additive noise. When $L(q)$ is positive one expects that additive
noise will only increase $\langle x_i^q\rangle$ further. When
$L(q)$ is negative $\langle x_i^q\rangle$ cannot be expected
to go to zero as even in the absence of any multiplicative force
$\langle x_i^q\rangle$ should reach a non-zero steady state
value. Therefore we expect that in the limit $t\rightarrow\infty$
$\langle x_i^q\rangle$ has a nonzero steady state value when
$L(q)$ is negative, and becomes ill defined  when $L(q)$ is 
positive.

Consider fig. \ref{fig:l(q)}. Here two examples of possible $L(q)$'s are
illustrated.
An important restriction to bear in mind is that
$L(q)$ is a convex function of $q$.  
The solid curve represents the case where $L(q)$ changes sign
as a function of $q$. $q^*$ represents the point where $L(q^*)~=~0$.
From the above argument, for $q > q^*$ the moment
$\langle x_i^q\rangle$ becomes ill defined in steady state.
Below this point the moments are well defined. Such behavior
implies that the PDF of $x$ has a power-law
tail $P(x) \propto x^{-q^*-1}$.

The other case of interest is where $L(q)$ never crosses through
zero for positive $q$. In this case all moments are defined and
$P(x)$ should have a tail that vanishes more quickly then any
power. In this case,  it does not seem possible
to obtain the actual form for the tail of $P(x)$ by the heuristic
considerations just presented. However
in section \ref{sec:nongaussian} a model will be analyzed
where the form of the tail
can be completely determined by a knowledge of $L(q)$ for large
$q$.

The probability distribution  with no additive noise,
eqn. (\ref{eq:plnphi}) can be
obtained from eqn. (\ref{eq:falpha}) with a knowledge of
$L(q)$. With the convex monotonically decreasing $L(q)$ just 
considered, it can be seen from these equations that $f(\alpha)$
is only defined up to a finite maximum value $\alpha_{max}$.
After this point, the probability distribution is zero. 
This means that $L(q)$'s of this type derive from highly
non-gaussian multiplicative noise $\gamma_i$ 
(cf. eqn. (\ref{eq:discrete})). The probability distribution
of $\gamma_i$ must be zero beyond a certain value for this 
kind of curve, $L(q)$.

\section{Gaussian case}
\label{sec:gaussian}
We will now solve for the averaged moments of $x (\tau )$. Since the
noise is stationary, if a convergent moment exists then $\langle  x^n (\tau ) \rangle  ~=~
\langle x^n (0) \rangle$.  Solving eqn. (\ref{eq:multiplicative}) for $x(0)$ gives
\begin{equation}
x(0) ~=~ \int_0^{\infty} e^{-\int_0^s (1- \gamma (\tau )) d \tau} \eta (s) ds
\end{equation}
The equation for the 2nth moment averaged over both $\gamma$ and $\eta$ is
\begin{equation}
\langle x^{2n}(0)\rangle  ~=~ \int_0^{\infty} \dots \int_0^{\infty}
\langle e^{-\sum_{i=1}^{2n} \int_0^{s_i}
(1-\gamma (\tau)) d\tau} \rangle  
\langle \eta (s_1) \eta (s_2) ... \eta (s_{2n}) \rangle  ds_1 \dots ds_{2n}
\end{equation}
Now we take $\eta$ to be to be gaussian white noise so that Wick's theorem can be applied to
the right hand side to separate
out the average into all possible combinations of two point correlation functions.
Each term gives the same contribution so one obtains
\begin{equation}
{(2n)! \over {2^n n! }} ({2T \over k})^n
\int_0^{\infty} \dots \int_0^{\infty} \langle e^{-\sum_{i=1}^{2n} \int_0^{s_i}
(1-\gamma (\tau)) d\tau} \rangle \delta (s_1 - s_{n+1})\cdots \delta (s_n -s_{2n})
 ds_1 \dots ds_{2n}
\label{eq:wick}
\end{equation}
Now the limits of integration can be restricted to $s_1 > s_2  \dots > s_n > 0$
at the expense of multiplying by $n!$,
as each possible ordering of the $s_i$'s gives an equal contribution. The average
with respect to $\gamma$ of the exponential can also be performed.  This gives
\begin{eqnarray}
\langle &x&^{2n}(0)\rangle ~=~ \nonumber \\
(2n)! & & ({T \over k})^n\int_0^\infty\int_0^{s_1}\cdots\int_0^{s_{n-1}}
e^{-2\sum_{i=1}^n s_i ~+~ 2\int\int S(s) \sigma (s-s') S(s') ds ds'}
ds_1\dots ds_n
\label{eq:general}
\end{eqnarray}
Here we use the Heaviside function $\theta (s)$ to write
\begin{equation}
S(s) ~=~ \sum_{i=1}^n \theta(s_i -s).
\end{equation}
This has the form of a staircase.
\subsection{White noise correlations}
\label{sec:white}
Before we proceed to analyze the case of general correlations in $\gamma$, we
examine the case where 
\begin{equation}
\sigma (\tau -\tau ' ) ~=~ \sigma _o \delta (\tau -\tau ' ).
\label{eq:shortrange}
\end{equation}
A differential equation for the time evolution of the probability, $P(x,t)$ can be
derived and has the form of a generalized Fokker-Planck equation~\cite{tatarskii}.
\begin{equation}
{\partial P \over \partial t} ~=~ {\partial\over\partial x} [x -{\sigma_o\over 2} x
+{\partial\over\partial x} ({\sigma_o\over 2}x^2+ {T\over k})] P
\end{equation}
The steady state solution corresponding to zero particle current, that is, no
particles being created or destroyed at the boundaries, is found by setting the
expression to the right of the ${\partial\over\partial x}$ equal to 0. Solving
this gives
\begin{equation}
P(x) \propto {1 \over {({2T\over k\sigma _o} + x^2 )^{{1\over \sigma _o}+{1\over 2}}}}
\label{eq:p(x)}
\end{equation}
A related three dimensional version of this has been derived in the context
of polymers in turbulent flow ~\cite{armstrong}. It shows that the large $x$
behavior of $P(x)$ has a power law tail $P(x) \sim x^{-1-2/\sigma _o}$. 
We can check that eqn. (\ref{eq:general}) is correct by substituting eqn. (\ref{eq:shortrange}). 
into it. One can then decouple all of the $s_i$ integrations by making
the change of variables 
\begin{equation}
\Delta _1 \equiv s_2 -s_1,~\Delta _2 \equiv s_3 - s_2, \dots 
\Delta _{n-1} \equiv s_n - s_{n-1},
\Delta _n \equiv s_n
\label{eq:change}
\end{equation}
So the right hand side of eqn. (\ref{eq:general}) becomes
\begin{equation}
{(2n)! \over {2^n }} ({2T \over k})^n\int_0^{\infty} \dots \int_0^{\infty}
e^{\sum_{i=1}^n  (-2i\Delta _i + 2 \sigma _o i^2 \Delta _i)} d\Delta_1\dots d\Delta_n
\end{equation}
Performing the intergration gives
\begin{equation}
\langle x^{2n}\rangle ~=~
{(2n)! \over {2^n }} ({T \over k})^n\prod_{i=1}^n{1 \over {1-\sigma _o}i}
\label{eq:shortmoment}
\end{equation}
This can be compared with the answer obtained from eqn. (\ref{eq:p(x)})
by computing moments. The integrals can be done in closed form ~\cite{crc}
and give the identical result.
\par
Clearly the precise form of the moments $\langle x^{2n}\rangle$, and hence $P(x)$, 
is dependent on the whole function $\sigma (s)$. However we shall now see that the
large $x$ behavior of $P(x)$ only depends on the total area under $\sigma (s)$. 
\subsection{General Asymptotic Behavior}
\label{sec:asym}
To determine the power law exponent of $P(x)$ for large $x$, for general
correlations $\sigma (s)$, one locates at which value of 
$n$ eqn.(\ref{eq:general}) becomes
divergent. That is if we define $\alpha$ so that $P(x) \propto x^{-\alpha}$ for large
$x$, then $\langle x^{2n}\rangle$ becomes divergent when $2n-\alpha ~=~ -1$ or
$\alpha ~=~ 2n +1$.
For example for short range correlations, eqn. (\ref{eq:shortmoment}),
becomes divergent when $n ~=~ 1/\sigma _o$. Therefore $\alpha = 1 + 2/\sigma _o$,
in agreement with eqn. (\ref{eq:p(x)}). Therefore the object of this section is to
locate the value of $n$ where eqn. (\ref{eq:general}) becomes divergent, for
general but normalizable $\sigma (s)$.
\par
To do this we rescale the $s_i$ in eqn. (\ref{eq:general})
\begin{equation}
s \equiv s_1
p_2 ~\equiv {s_2 \over s_1},~p_3 \equiv {s_3 \over s_1},\dots 
~p_n \equiv {s_n\over s_1}
\end{equation}
so that
\begin{eqnarray}
\langle &x&^{2n}\rangle ~=~
(2n)! ({T\over k})^n \int_0^\infty s^{n-1} F(s) ds \nonumber \\
F(s) \equiv \int_0^1\int_0^{p_3}\dots\int_0^{p_n}
&e&^{-2s\int R(p) dp +2s^2 \int\int R(p)\sigma(s(p-p')) R(p' )dpdp'}dp_2\dots dp_n\\
&R&(p) ~=~ \theta (1-p) +\sum_{i=2}^n \theta (p_i -p) \nonumber
\end{eqnarray}
The divergence in $\langle x^{2n}\rangle$ is therefore controlled by the large $s$ behavior
of $F(s)$. For large $s$ the function $\sigma (s(p-p'))$ becomes very sharply
peaked. One can define a limit of functions approaching a $\delta$ function by
\begin{eqnarray}
\lim_{s\rightarrow \infty} \sigma(s(p-p')) ~=~ {\delta (p-p')\over s} \sigma _o \\
\sigma _o ~=~ \int_{-\infty}^{\infty} \sigma (p) dp
\end{eqnarray}
For large $s$, the leading order term in the exponent of the integrand is identical
to the short range case
\begin{equation}
-2s[\int R(p) dp -  
\sigma_0 \int\int R(p)\delta (p-p') R(p' )dpdp' + {\cal O}(1/s^P)]
\end{equation}
Where the exponent $P > 0$ depends on the form of $\sigma (s)$.
As $n$ increases, the staircase function $R(p)$ becomes larger. Since the second term
is quadratic, and the first term, linear in $R$, for large enough $n$ ithe
expression inside the square brackets becomes positive so that the entire
expression will increase linear with s, for some values 
of the $p_i$'s. This leads to a divergence.
Lower order corrections in $s$, cannot change this point of divergence as it 
is determined by the sign of the slope of this expression for large s.
\par
Therefore for large $s$, $F(s)$ will only depend on $\sigma _o$,
the total area under $\sigma (s)$.

\section{short range non-gaussian noise}
\label{sec:nongaussian}

We now relax the restriction that the multiplicative noise be Gaussian,
but restrict the analysis to the case of short range correlations. 
In section \ref{sec:absence} eqn.(\ref{eq:multiplicative}) was
discretized.  With additive noise included it can also be
solved in discretized form, but the results are very similar looking
to the continuous case. We can, with little loss of generality use
the continuous form of the equation to analyze this problem.
With continuous notation the equation for the Lyapunov exponents
read
\begin{equation}
\langle e^{q\int_0^t\gamma(\tau)d\tau}\rangle ~=~ e^{L(q) t}
\label{eq:l(q)}
\end{equation}
Defining $D \equiv 2T/k$ we can still salvage the previous calculation
up to eqn. (\ref{eq:wick}), with $(\gamma(\tau) -1) \rightarrow \gamma(\tau)$, 
and have
\begin{equation}
\langle x(0)^{2n}\rangle ~=~
{(2n)! \over {2^n n!}} D^n\int_0^{\infty} \dots \int_0^{\infty}
\langle e^{2\sum_{i=1}^n \int_0^{s_i}\gamma (\tau )d\tau}\rangle 
ds_1\dots ds_n .
\end{equation}
Using the change of variables of eqn. (\ref{eq:change}) and also 
eqn. (\ref{eq:l(q)}) defining the Lyapunov exponents, this yields
\begin{equation}
\langle x(0)^{2n}\rangle ~=~
{(2n)! \over {2^n }} D^n \int_0^\infty e^{\sum_{i=1}^n L(2i)\Delta_i}
d\Delta_1\dots d\Delta_n
\end{equation}
Performing the integrations over the $\Delta_i$'s gives the
final answer
\begin{equation}
\langle x(0)^{2n}\rangle ~=~ (2n)! {({D\over 2})}^n \prod_{i=1}^n
{1\over {-L(2i)}}
\label{eq:mom(l)}
\end{equation}
The above analysis has assumed that the discretization time step
$\Delta t$ is very small so that $L(q)\Delta t$ is small. Similar
expressions to above can be derived for the discretized case without
this restriction on $\Delta t$ which amounts to replacing the last 
integration over the $\Delta$'s by a discrete summation. This gives
\begin{equation}
\langle x(0)^{2n}\rangle ~=~ 
(2n)! ({D\over 2})^n \prod_{i=1}^n
{{\Delta t}\over {1-e^{L(2i)\Delta t}}}
\end{equation}

\subsection{Relation between the distribution and L(q)}

By knowing the moments of $x$ one can obtain its PDF. 
Now that we have an expression for the
moments eqn. (\ref{eq:mom(l)}) as a function of 
the Lyapunov exponents $L(q)$, we can
consider the problem of how different forms of  $L(q)$ affect
the probability distribution of $x$.  Different forms of $L(q)$
were shown in fig. \ref{fig:l(q)} and discussed in 
section \ref{sec:heuristic}. The solid line in this figure
represents the case of finite $q^*$, the point beyond which the
Lyapunov exponents become positive.
In this case eqn. (\ref{eq:mom(l)}) implies that for $2n > q^*$
the moments are not defined, for $q^*$ an integer. This implies
that the p.d.f. has a power law tail $P(x) dx ~= x^{-q^*-1}$.
This is in agreement with the results found earlier for Gaussian
multiplicative noise, and the heuristic 
argument of section \ref{sec:heuristic}. 

The other case to consider is when $L(q)$ is always negative for
positive $q$.  Consider the case where the asymptotic form of
$L(q)$ for large q is proportional to $-q^\beta$. The restriction
of convexity implies $0\leq \beta  \leq 1$. To find the tail of
$P(x)$, one starts by substituting
$L(2i) ~=~ -K i^\beta$ in eqn. (\ref{eq:mom(l)}) and uses
Stirlings approximation to simplify the expression. 
\begin{equation}
\langle x^{q}\rangle 
~=~ e^{q(\ln(q)-1)-\beta (q/2)(\ln(q/2)-1) +Cq}
\end{equation}
where $C$ is a constant whose value will not be important
in the conclusions to this analysis.
The asymptotic form of $P(x)$ can be obtained by a steepest
descent analysis. In general
\begin{equation}
\exp (D(q)) \equiv \langle x^{q}\rangle ~=~ \int P(x) x^q dx
\end{equation}
Changing variables in the last integration to $u = \ln x$
and performing steepest descent, which is valid for large $q$,
one has $D(q) = f(u(q))+qu(q)$. Here $\exp (f(\ln x)) d(\ln x)
= P(x) dx$ and
$u=u(q)$ is determined by the saddle point equation $df/du = -q$.
We want to solve
for $f(u)$ which can be done by Legendre transform to a different
``ensemble'' directly borrowing this method from thermodynamics.
Then $f(u) = D(q) - qu$ and $dD/dq ~=~ u$. Carrying out out the
algebra yields the result
\begin{equation}
P(x ) \propto e^{-Kx^{2/(2-\beta )}}
\end{equation}
where $K$ is a constant. This is stretched exponential
behavior. An application of this to experimental systems
will be discussed in the following paper~\cite{many}.
Note that when $\beta \rightarrow 0$, the tail becomes 
exponential. The limit where $L(q)$ approaches a constant
for large $q$, corresponding to $\beta ~=~ 0$, has qualitatively
different behavior for $f(\alpha )$ defined in eqn. \ref{eq:plnphi}
for finite $\beta$. In the former case $f(\alpha)$ goes to 
negative infinity at a finite value $\alpha = \alpha_{max}$,
whereas in the latter case $f(\alpha )$ goes to a finite
value at $\alpha = \alpha_{max}$.

\section{Conclusions}

The PDF for eqn. (\ref{eq:multiplicative}) has been analyzed in one
dimension under a variety of conditions,
long range time correlations with gaussian statistics,
and nongaussian noise with short range temporal correlations.
It has been possible to classify the steady state PDF's with additive
noise by examining the Lyapunov exponents. When the Lyapunov
exponents $L(q)$ pass through zero at finite $q$, the tails are
power law. The exact exponent was worked out for 
Gaussian multiplicative noise and was found to be independent
of the strength of the additive noise.
If the Lyapunov exponents remain negative for all positive $q$ it
was argued that the PDF, P(x) should be of the form 
$\log P(x)\propto -x^p$ where $p$ is a power between $0$ and $2$. 

It is interesting to note that many of the conclusions here hold for 
coupled many component equations of the same form. This will be
seen in the following paper.

\acknowledgments
The author thanks Herbert Levine, Eliot Dresselhaus, Ken Oetzel,
and Matthew Fisher for useful discussions. This research
was supported by the NSF under grant DMR-9112767.

\begin{figure}
\caption{ Examples of two different $L(q)$'s. The solid line represents
an  $L(q)$ that intersects the horizontal axis at finite $q$. The
dashed line represents one that does not. 
\label{fig:l(q)}}
\end{figure}
\newpage
\begin{figure}[ht]
\centering
\leavevmode
\epsfysize=10 cm \epsfbox{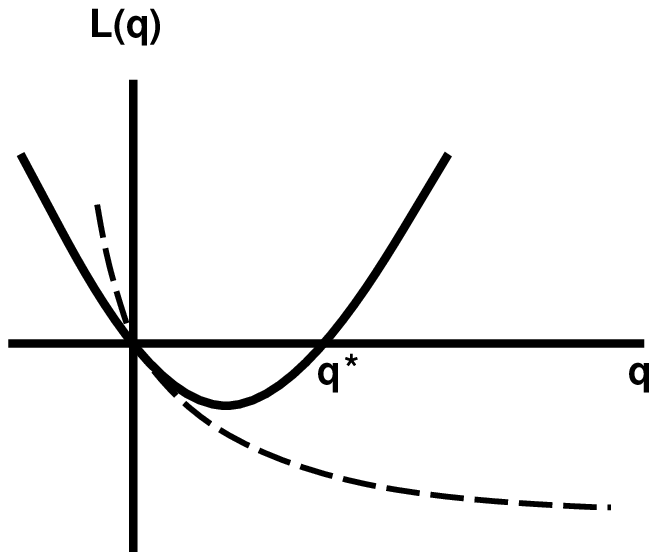}
\end{figure}
\vskip 3in
\centerline{Fig. 1}
\end{document}